\documentclass[pra,twocolumn,showpacs,groupedaddress,notitlepage,superscriptaddress,floatfix,nofootinbib]{revtex4-2}
\UseRawInputEncoding
\usepackage{subfigure}
\usepackage{braket}
\usepackage{tikz}
\usepackage{amsmath}
\usepackage{listings}
\usepackage{algorithmicx}
\usepackage[ruled]{algorithm} 
\usepackage{algpseudocode}
\usepackage{amsfonts}
\usepackage{float,mathdots}
\usepackage{amssymb}
\usepackage{accents}
\usepackage{amsthm}
\usepackage{hyperref}
\usepackage{blkarray}
\usepackage{svg}
\usepackage{booktabs}

\newcommand\bigDiamond{\mathop{\mathpalette\bigDiamond\relax}}

\renewcommand{\[}{\begin{equation}}
\renewcommand{\]}{\end{equation}}

\usepackage{bbm}
\usepackage{appendix}

\hypersetup{
    colorlinks=true,   
    linkcolor=cyan,    
    citecolor=magenta, 
    filecolor=magenta, 
    urlcolor=cyan,     
    runcolor=cyan
}
\usepackage[capitalise]{cleveref} 

\newcommand{\cyan}[1]{\textcolor{cyan}{#1}}

\newcommand{\DL}[1]{\cyan{[DL: #1]}}
\renewcommand{\DL}[1]{}

\begin{document}

\title{Optimizing continuous-time quantum error correction for Markovian and non-Markovian noise models}
\author{Anirudh Lanka}
\affiliation{Department of Electrical \& Computer Engineering, University of Southern California, Los Angeles, California}

\author{Shashank Hegde}
\affiliation{Department of Electrical \& Computer Engineering, University of Southern California, Los Angeles, California}

\author{Todd A. Brun}
\affiliation{Department of Electrical \& Computer Engineering, University of Southern California, Los Angeles, California}

\begin{abstract}
We present a protocol using neural networks to simultaneously optimize the quantum error-correcting code space and the corresponding recovery map in the framework of continuous-time quantum error correction. Given a Hilbert space and a noise process---potentially correlated across both space and time---the protocol identifies the optimal recovery strategy, measured by the average logical state fidelity. This approach enables the discovery of recovery schemes tailored to arbitrary device-level noise.
\end{abstract}

\maketitle

\section{Introduction}

Quantum systems are always prone to environmental interactions leading to decoherence. To mitigate this, quantum information is often encoded in a higher dimensional system so that the redundancy can make the system resilient to certain errors---the procedure is known as quantum error correction (QEC). Among various choices of quantum error correcting codes, stabilizer codes are particularly appealing due to their ease of description using Pauli operators and having close analogies to classical error correction schemes. However, quantum noise affecting physical qubits can exhibit rich and complex dynamics: it may be non-Markovian, cause leakage out of the computational subspace, introduce non-Pauli errors, or combinations of these effects.  In such cases, standard stabilizer codes may be suboptimal. This motivates the need for channel-adapted recovery schemes.

Channel-adapted recovery scheme, as the name suggests, aims to provide measurement and unitary correction operators for a given noise channel that are \emph{optimal} for a chosen metric (such as the decoherence rate or the average infidelity with respect to the desired code state). Ref. \cite{leung_approximate_1997} showed that when the noise is not assumed to cause single-qubit Pauli errors, but cause amplitude damping errors (as often can be), $4$ physical qubits are sufficient to protect quantum information reliably (as opposed to $5$ qubits required by the perfect $[\![5, 1, 3]\!]$ code). This motivated the development of various optimization techniques to design the encoding and the recovery strategies for a given noise model \cite{petz_optim_cond, reimpell_iterative_2005, fletcher_channel-adapted_2007, fletcher_channel-adapted_2008, kosut_robust_2008, kosut_quantum_2009, taghavi_channel-optimized_2010, convy_machine_2022, wang_automated_2022, zeng_approximate_2023, zheng_near-optimal_2024, dutta_noise-adapted_2025}.

The standard implementation of QEC---whether using stabilizer or channel-adapted codes---relies on performing projective measurements to extract error syndromes, followed by unitary corrections to restore the state to the code space. This discrete model assumes that recovery operations are fast relative to the noise time-scale. In practice, however, the noise is perpetual and the environmental interactions continue to introduce errors. Consequently, QEC should be implemented in repeated cycles. As the interval between these cycles approaches zero, one enters a regime where noise and recovery occur effectively simultaneously. This continuum limit forms the theoretical foundation of continuous-time quantum error correction (CT-QEC) \cite{paz_continuous_1998, ahn_continuous_2002, sarovar_practical_2004, brun_continuous-time_2013, ippoliti_perturbative_2015, chase_efficient_2008, chen_continuous_2020, convy_machine_2022}.

In the framework of CT-QEC, the system is continuously subjected to both \emph{weak} noise and \emph{weak} recovery operations in a \emph{small} time interval. A crucial aspect of CT-QEC is the choice of measurement observables. Unlike in discrete QEC---where many different choices of measurements and correction operators lead to equivalent performance---in the continuous-time setting, weak versions of different measurements and corrections may perform differently, even if their strong versions are equivalent for discrete QEC. In particular, weakly measuring stabilizers and applying weak Pauli corrections may result in suboptimal performance \cite{oreshkov_continuous_2007, hsu_method_2016}. For larger codes, identifying the optimal measurement basis becomes considerably less straightforward.

In this work, we combine the idea of channel-adapted recovery scheme to identify the optimal basis to perform CT-QEC for a given noise channel. By construction, we allow the noise channel to be an arbitrary completely-positive and trace-preserving (CPTP) map that can allow correlations in both space and time. We perform the optimization using neural networks to find the (locally) optimal recovery channel and the code space. The set of all CPTP maps is convex---this means that standard gradient approaches can be utilized to perform a single backward propagation step. However, the set of all subspaces (which may act as the code space) forms the Grassmannian manifold. We use a Riemannian gradient descent technique to find the optimal code space. Because the
Riemannian gradient can be obtained by projecting the Euclidean gradient onto the tangent space of the Grassmannian, the gradients for the code space and recovery channel can be computed from the same backward pass and updated simultaneously. This differs computationally from existing approaches that alternate between optimizing the encoding and recovery channels. The logical performance, however, is fundamentally a property of the matched code--recovery pair rather than of either component in isolation. Let
\[
    \varepsilon(\mathcal{C},\mathcal{R})
\]
denote the logical infidelity obtained from code space $\mathcal{C}$ and recovery channel $\mathcal{R}$. The joint optimum satisfies
\[
    \varepsilon_{\mathrm{joint}} = \min_{\mathcal{C},\mathcal{R}} \varepsilon(\mathcal{C},\mathcal{R}) \leq \min_{\mathcal{R}} \varepsilon(\mathcal{C}_0,\mathcal{R}),
\]
for any fixed code $\mathcal{C}_0$, and
\[
    \varepsilon_{\mathrm{joint}} \leq \min_{\mathcal{C}} \varepsilon(\mathcal{C},\mathcal{R}_0),
\]
for any fixed recovery $\mathcal{R}_0$. These inequalities can be strict because the two optimization variables are coupled. For a fixed code, distinct physical errors may produce indistinguishable corrupted code subspaces, in which case no recovery can perfectly correct all of them. Conversely, a recovery channel is constructed to measure and map the noise-induced subspaces of a particular code; after the code space is changed, that same recovery is generally no longer matched to the new error subspaces. Thus a code cannot, in general, be assigned a
decoder-independent logical performance. We provide an example in this work that illustrates this behavior: adapting the recovery for a fixed code improves its performance, but the jointly optimized code--recovery pair performs better because the fixed code contains syndrome collisions that no recovery can eliminate. Here, ``simultaneous'' refers to computing and updating both sets of parameters within one joint optimization step; we do not claim that simultaneous gradient updates must outperform an ideally converged alternating method that also optimizes both variables.

CT-QEC is particularly appealing due to its ability to suppress the effects of non-Markovian noise \cite{oreshkov_continuous_2007, chen_continuous_2020, nila2025continuousquantumcorrectionmarkovian}. This is because the continuous measurements (that can be weak) mitigates the system-bath correlations that normally build up over the bath's memory time. This makes the dynamics effectively Markovian and pushes the state toward a syndrome space through the quantum Zeno effect. This is analogous to dynamical decoupling \cite{dd_zeno}: just as rapid control pulses average the system-bath interaction and filter out low-frequency environmental noise, frequent measurements truncate the bath's memory kernel and repeatedly push the system toward the syndrome space. In both cases the suppression arises from imposing a fast external timescale that outpaces the bath dynamics, but in CT-QEC this is achieved through continuous measurements rather than unitary control pulses; the Zeno effect plays the role that time-domain averaging plays in dynamical decoupling. In this work, we exploit this regime to design codes and recovery schemes for noise channels that show non-Markovian system behavior.

Another important aspect of CT-QEC is how measurement outcomes are processed. If the full measurement record is used, the resulting feedback is classified as indirect; if only the most recent measurement outcome is used to determine the next unitary rotation, the feedback is considered direct. While in principle, it may be beneficial to process the entire measurement record, doing so in real time may become computationally expensive. Moreover, it turns out that the improvement we can gain by processing the record to keep track of the actual states is small \cite{brun_continuous-time_2013}. Therefore, we choose to use only the latest measurement outcome to perform the weak unitary correction in a small interval.

\subsection{Continuous-time Quantum Error Correction} \label{sec:ct_qec}

The dynamics of a system subject to continuous-time quantum error correction can be written as
\begin{equation}\label{eq:general_ct_qec}
    \dot \rho = (\mathcal{D}_N + \mathcal{D}_R)(\rho),
\end{equation}
where $\mathcal{D}_{N}$ ($\mathcal{D}_{R}$) are the traceless super-operators describing the dynamics of the noise (recovery). The goal of CTQEC is to preserve $\rho$ for as long as possible. We consider a simple example of protecting a single qubit in the $\ket 0$ state against both Markovian and non-Markovian bit-flip errors. While this system does not actually store any information, it serves as a useful example to understand important features of CTQEC. This single qubit can be regarded as a stabilizer code, where the code space is spanned by $\ket{0}$ and its stabilizer is $Z$.

\subsection{Impact of measurement basis}

In this section, we show why the choice of the measurement basis impacts the overall performance of CT-QEC. Suppose the noise $\mathcal D_N$ causes Markovian system errors. We model this behavior through Lindblad evolution. Suppose the errors happen as bit-flips at a rate $\gamma$. The Lindbladian $A$ can be represented by $\sqrt{\gamma}X$, where $X$ denotes the Pauli-$X$ operator. To perform a discrete strong correction, one can measure the stabilizer $Z$ and apply a unitary correction $X$ if the measurement outcome is $-1$. One heuristic approach to convert this into a continuous-time procedure is given in \cite{ahn_continuous_2002}: weakly measure $Z$, and rotate about $X$ based on the \emph{entire} measurement history. The dynamics of the state can be represented by the stochastic master equation:
\[\label{eq:z_x_map}
\begin{split}
    \mathrm{d}\rho &= -i[H_F(t), \rho]\,\mathrm{d}t + \gamma\mathcal{D}[X]\rho\,\mathrm{d}t + \\
    &\qquad\qquad \kappa_{Z}\mathcal{D}[Z]\rho\,\mathrm{d}t +\sqrt{\kappa_{Z}}\mathcal{H}[Z]\rho\,\mathrm{d}W_t,
\end{split}
\]
where the superoperators $\mathcal{D}$ and $\mathcal{H}$ are defined as
\begin{subequations}
    \begin{align}
        \mathcal{D}[A]\rho &\equiv A\rho A^\dagger - \frac{1}{2}\{A^\dagger A, \rho\} \\
        \mathcal{H}[A]\rho &\equiv A\rho + \rho A^\dagger - \langle A + A^\dagger \rangle \rho,
    \end{align}
\end{subequations}
$H_F(t)$ is the feedback Hamiltonian that depends on the measurement record, $\kappa_{Z}$ is the measurement rate, and $\mathrm{d}W_t \sim \mathcal N(0, \mathrm{d}t)$ denotes the Wiener increment. The measurement current reads
\[
    J(t) = \braket{Z}_{\rho(t)} + \frac{1}{2\sqrt{\kappa_{Z}}}\frac{\mathrm{d}W_t}{\mathrm{d}t}.
\]
To construct the feedback Hamiltonian, we first estimate the density matrix $\hat{\rho}(t)$ by simulating the master equation
\[
\begin{split}
    \mathrm{d}\hat{\rho} &= -i[H_F(t), \hat{\rho}]\,\mathrm{d}t + \gamma \mathcal{D}[X]\hat{\rho}\,\mathrm{d}t + \kappa_{Z}\mathcal{D}[Z]\hat{\rho}\,\mathrm{d}t + \\
    &\qquad\qquad 2\kappa_{Z}\mathcal{H}[Z]\hat{\rho}(J(t) - \braket{Z}_{\hat{\rho}(t)})\,\mathrm{d}t.
\end{split}
\]
From the density matrix estimate, we can construct the feedback Hamiltonian with the control parameter $\omega$:
\[
    H_F(t) = \omega\;\text{sgn}\braket{Y}_{\hat{\rho}(t)} X,
\]
where $\text{sgn}$ denotes the sign function, and $\omega$ is the maximum feedback strength that can be applied: it is a \emph{bang-bang} control scheme, meaning that the control parameter $\omega$ is always at the maximum or minimum value possible. The measurement rate and the maximum feedback strength are chosen to be $\kappa_Z = 64\gamma$, and $\omega=128\gamma$. This is an example of indirect feedback protocol. It is immediately clear that when the system size grows, keeping track of the expectation values becomes exponentially harder.

However, this is not the only way to perform CT-QEC. For instance, one could also weakly measure $X$ and rotate about $Y$ based only on the \emph{latest} measurement outcome \cite{paz_continuous_1998,oreshkov_continuous_2007, hsu_method_2016}. The channel corresponding to such a recovery can be obtained by taking the convex combination of the identity channel and the strong recovery channel:
\[
    \tilde{\mathcal{R}}_{\text{X}} = (1 - \epsilon^2)\mathcal{I} + \epsilon^2 \mathcal{R}_{X}.
\]
Here $\epsilon=\sqrt{\kappa_X\delta t}$, $\kappa_{X}$ is the \emph{recovery} rate and $\delta t$ is an arbitrary small time interval. The strong recovery channel has Kraus operators $R_0 = |0\rangle\langle0|$, and $R_1 = |0\rangle\langle 1|$.

\begin{figure}
    \centering
    \includegraphics[width=\linewidth]{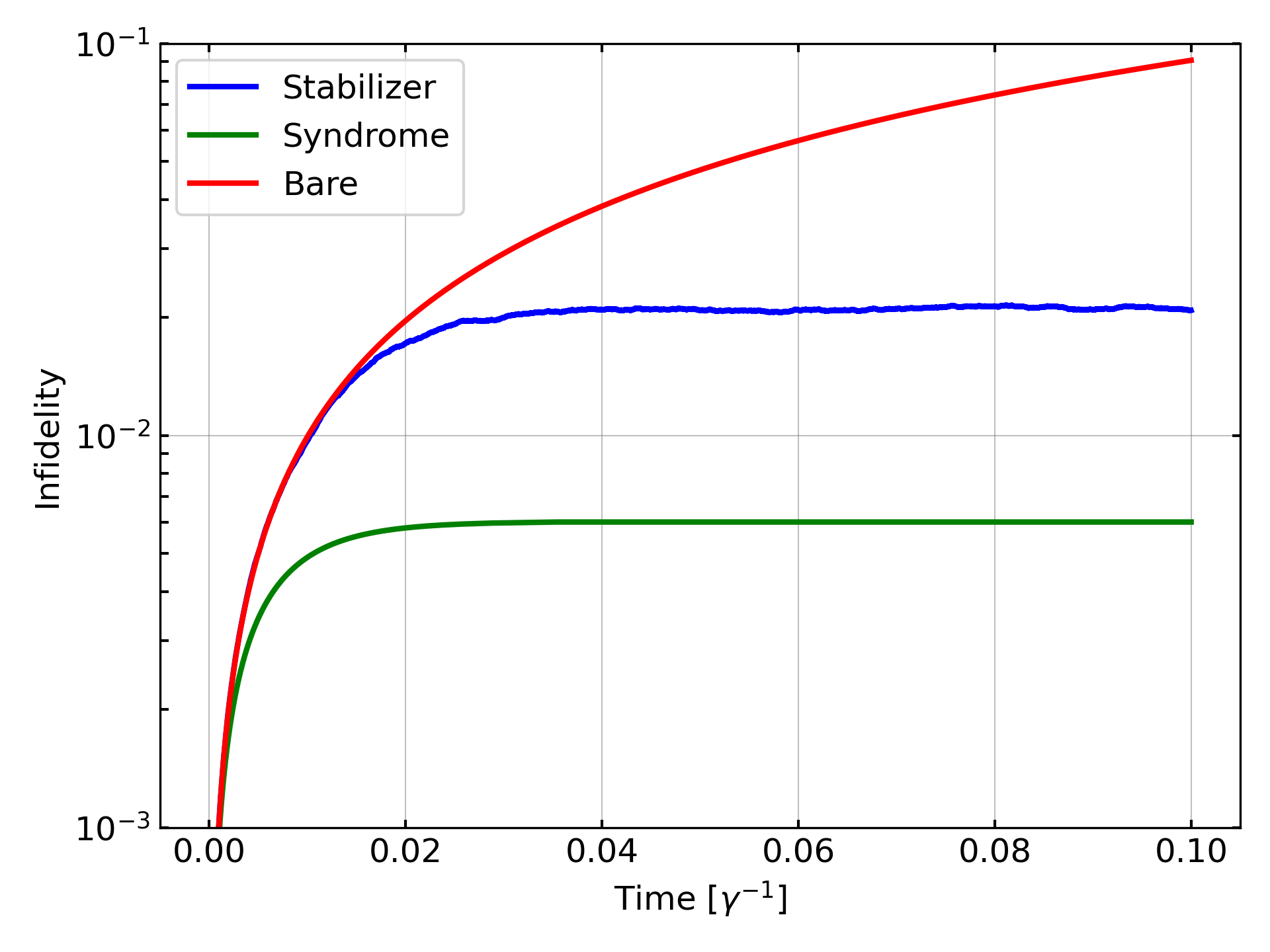}
    \caption{Infidelity as a function of time under the effect of Markovian bit-flip noise and continuous measurements in the $Z$ ($X$) basis and unitary rotation around $X$ ($Y$) axis. The ensemble average is taken over $10^5$ trajectories.}
    \label{fig:markov_1q}
\end{figure}

\begin{figure*}
    \centering
    \includegraphics[width=0.8\linewidth]{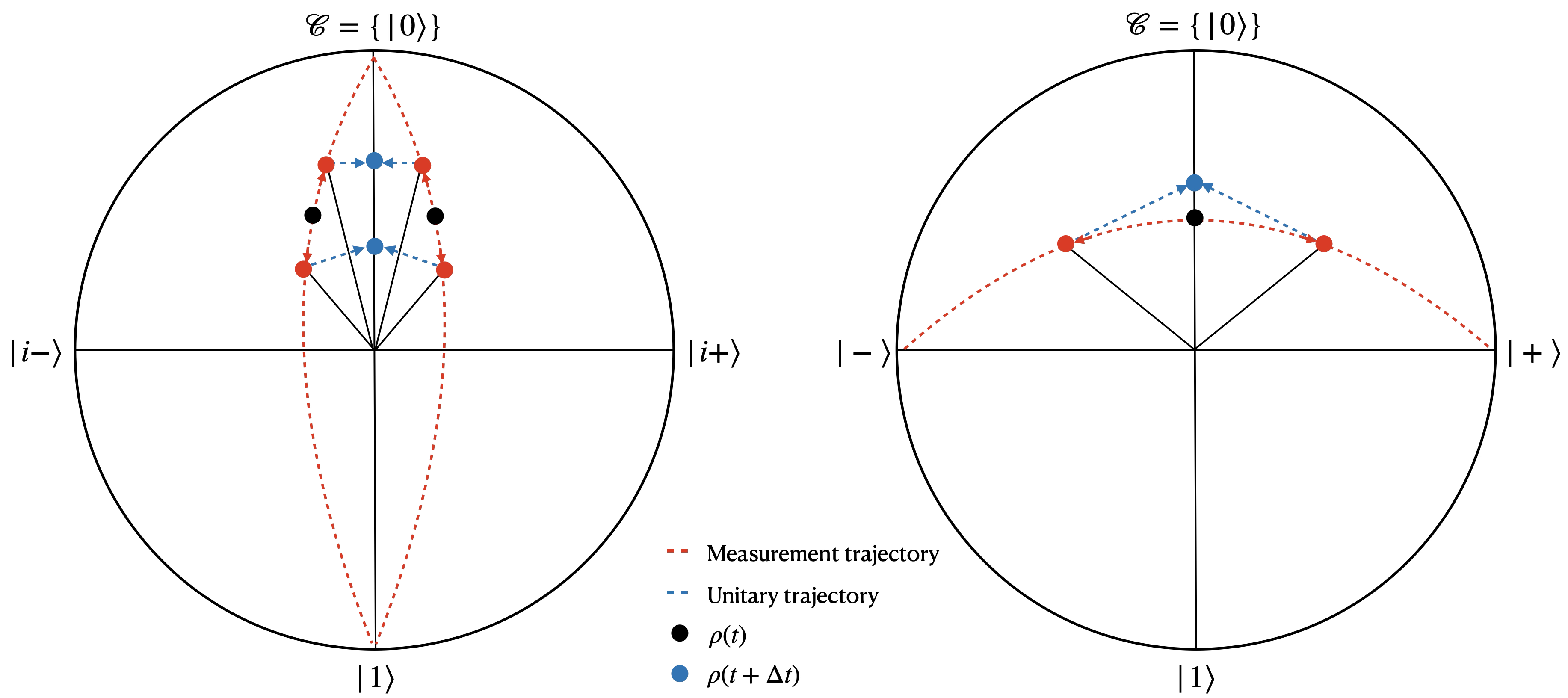}
    \caption{(Left) Continuously measure in the $Z-$basis, and apply unitary rotations about the $X-$axis. The state does not move towards the target code state on average. (Right) Continuously measure in the $X-$basis, and apply unitary rotations about the $Y-$axis. On average, the purity of the state increases, and it reaches the north pole faster.}
    \label{fig:ctqec_meas_basis}
\end{figure*}

To compare the performance between different protocols, we should adjust the parameters in both protocols so that the comparison is ``fair'' in the sense that the strengths of the correcting maps are comparable. We measure the strength of the maps using the diamond norm, denoted by $\mathbb{D}[\cdot]$:
\[
    \mathbb{D}[\Phi] \equiv ||\Phi - \mathcal{I} ||_{\diamond}.
\]
Suppose $\tilde{\mathcal{R}}_{\text{Z}}$ denotes the ensemble averaged map corresponding to continuous measurements of $Z$ and a unitary rotation about $X$:
\[
    \tilde{\mathcal R}_{\text{Z}}(\rho) = \rho -i[\omega X, \rho]\delta t + \kappa_{Z} (Z \rho Z - \rho)\delta t.
\]
It turns out that the diamond norm is invariant of the sign of the feedback Hamiltonian. We find that
\[\label{eq:dnorm_zx}
    \mathbb{D}[\tilde{\mathcal R}_{\text{Z}}] = 2 \times 2.5615 \times \kappa_{Z} \delta t.
\]
On the other hand, the diamond norm for $\tilde{\mathcal R}_{X}$ is given by
\[\label{eq:dnorm_xy}
    \mathbb{D}[\tilde{\mathcal R}_{\text{X}}] = 2 \times \kappa_{X} \delta t.
\]
Comparing \cref{eq:dnorm_zx} and \cref{eq:dnorm_xy},
\[
    \kappa_{X} = 2.5615 \times \kappa_{Z}.
\]
We compare and plot the infidelities from both the strategies along with pure noise dynamics in \cref{fig:markov_1q}. The difference in performance can be explained intuitively using the Bloch sphere. Under the action of Markovian bit-flip noise, the state evolves towards the maximally mixed state. The effect of weak measurement in the $Z$-basis is to move the state slightly towards or away from the north pole ($\ket0$), depending on the measurement outcome. In the latter case, however, the state moves \emph{away} from the target state, and weak rotation about the $X$-axis maintains the length of the Bloch vector, so the purity of the state \emph{on average} does not increase. On the other hand, the channel corresponding to a weak measurement along the $X$-axis and unitary rotations about the $Y$-axis is non-unital---it increases the average purity. This is illustrated in \cref{fig:ctqec_meas_basis}.
In fact, Ref.~\cite{paz_continuous_1998} showed that the optimal way to prepare the $|0\rangle$ state is to measure in a basis perpendicular to the $Z$-axis. So, in the limit of weak recovery, the choice of the measurements and the corresponding correction Hamiltonian play a significant role in the overall performance \cite{oreshkov_continuous_2007}. For larger codes, this poses a natural optimization problem for the choice of the weak map and the code itself.

\subsection{Non-Markovian noise suppression}

In this section, we analytically show the suppression of non-Markovian noise through the action of frequent measurements. As before, we consider a single-qubit model, although the effects are analogous for larger codes. Suppose the noise is modeled through constant stochastic Hamiltonian
\[
    H_\text{err} = \lambda X,
\]
where $\lambda \sim U(-\gamma, \gamma)$ is uniformly distributed around $0$ by the error rate $\gamma$. This corresponds to a noise model with perfectly correlated (i.e., time-independent) fluctuations. One can verify that the associated spectral density is a delta function, with strong temporal correlations. The associated unitary evolution operator is 
\[
    U_\text{err}(t) = e^{-iH_\text{err}t}.
\]
The evolved state purely under the influence of the noise is
\[
    \rho(t) = U_\text{err}(t) \rho(0) U^\dagger_\text{err}(t).
\]
Assuming that the state is initialized in the pure $\ket 0$ state, the fidelity with respect to $\rho(t)$ is
\[
    \mathcal F_t = \bra{0} \rho(t) \ket{0} = \cos^2(\lambda t).
\]
The average fidelity over all noise realizations is
\[\label{eq:non_markov_noise_1q}
    \mathbb{E}_{\lambda}[\mathcal F_t] = \frac{1 + \text{sinc}(2\gamma t)}{2}.
\]

\begin{figure}
    \centering
    \includegraphics[width=\linewidth]{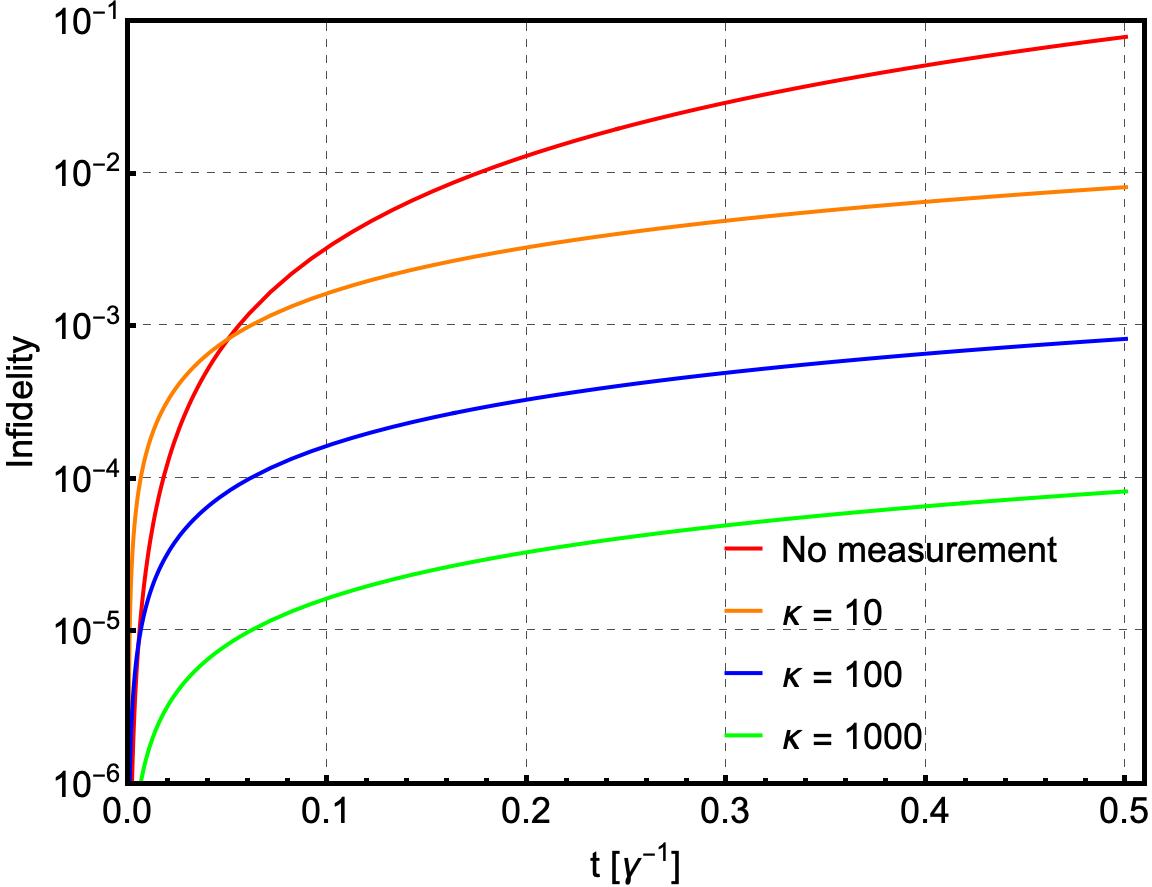}
    \caption{The Zeno effect suppresses the effects of non-Markovian noise. As the measurement rate $\kappa$ is increased, the average infidelity with the initial code state decreases.}
    \label{fig:zeno_1q}
\end{figure}

We now supplement this setup with continuous measurements of Pauli-$Z$ at a rate $\kappa$. The system dynamics can be represented by the stochastic master equation:
\[
    \mathrm{d}\rho = -i [H_\text{err}, \rho]\,\mathrm{d}t + \kappa \mathcal{D}[Z]\rho \,\mathrm{d}t +\sqrt{\kappa\eta}\mathcal{H}[Z]\rho\,\mathrm{d}W_t,
\]
where $\mathrm{d}W_t \sim \mathcal N(0, \mathrm{d}t)$ denotes the Wiener increment, and $\eta \in [0, 1]$ denotes the measurement efficiency. The Zeno effect also applies when we measure the system and ``forget'' the outcomes, i.e., $\eta = 0$. However, utilizing the measurement outcome and performing unitary corrections should, in principle, improve the performance. For the purpose of showing the suppression due to the Zeno effect, we chose to forget the outcomes for simplicity; the master equation becomes
\[
    \mathrm{d}\rho = -i [H_\text{err}, \rho]\,\mathrm{d}t + \kappa \mathcal{D}[Z]\rho\,\mathrm{d}t.
\]
The dynamics of the Bloch vector elements follow
\[
\begin{split}
    \dot{\braket{X}} &= 0 \\
    \dot{\braket{Y}} &= -2\lambda \braket{Z} - 2\kappa \braket{Y} \\
    \dot{\braket{Z}} &= 2\lambda \braket{Y}.
\end{split}
\]
In a more convenient matrix representation with the vector $\textbf{v} = (\braket{Y}, \braket{Z})^T$, the dynamics can be written as
\[\label{eq:non_markov_1q}
    \dot{\textbf{v}} = M \textbf{v},
\]
where the non-Hermitian matrix $M$ is
\[
    M = \begin{pmatrix}
        -2\kappa & -2\lambda \\
        2\lambda & 0
    \end{pmatrix}.
\]
Solving \cref{eq:non_markov_1q} yields
\[
    \begin{split}
        \braket{Z}_{\rho(t)} &= e^{-\kappa t} \bigg( \cosh\big(t\sqrt{\kappa^2-4\lambda^2}\big) + \\ 
        &\qquad\qquad\frac{\kappa}{\sqrt{\kappa^2-4\lambda^2}} \sinh\big(t\sqrt{\kappa^2-4\lambda^2}\big) \bigg).
    \end{split}
\]
Since we operate at a regime where the measurement rate is higher than the error rate, we can expand the above equation up to perturbative orders in $\lambda/\kappa$:
\[
    \braket{Z}_{\rho(t)} = \exp\bigg[-2 \bigg(\frac{\lambda^2}{\kappa}\bigg) t\bigg] + \mathcal{O}((\lambda/\kappa)^2),
\]
The average code state fidelity is given by
\[\label{eq:non_markov_noise_meas_1q}
\begin{split}
    \mathbb{E}_\lambda[\mathcal F^{\text{meas}}_t] &= \frac{1 + \mathbb{E}_\lambda[\braket{Z}_{\rho(t)}]}{2} \\
    &= \frac{1}{2}\left[1 + \frac{\sqrt{\pi}}{2}\frac{\text{Erf}(\sqrt{\Gamma t})}{\sqrt{\Gamma t}}\right] + \mathcal{O}((\gamma/\kappa)^2),
\end{split}
\]
where $\text{Erf}$ is the error function and the rate $\Gamma=2\gamma^2/\kappa$. We plot and compare the fidelities obtained from \cref{eq:non_markov_noise_1q} and \cref{eq:non_markov_noise_meas_1q} in \cref{fig:zeno_1q}. We observe that increasing the measurement rate suppresses the influence of non-Markovian noise. In a later section, we construct a code and corresponding recovery operators that mitigate the effects of non-Markovian noise on multi-qubit systems.

\subsection{Plan of this paper}

The rest of the paper is organized as follows. We begin by constructing a neural network architecture for optimizing the weak recovery channel and the code space for CT-QEC in \cref{sec:nn_arch}. In particular, we characterize the weak recovery channels with sets of arbitrary complex matrices in \cref{sec:weak_channel_char}; then we characterize the code space in \cref{sec:cs_char}. We then discuss the cost function when the noise is Markovian and non-Markovian in \cref{sec:cost}. We provide examples that show an improvement over conventional stabilizer codes for various noise processes in \cref{sec:examples}. We give concluding remarks and discuss the future scope of this work in \cref{sec:discussion}.

\section{Neural Network Architecture}\label{sec:nn_arch}

Trainable parameters in our model consist of (i) a weak recovery channel and (ii) a code subspace. We parametrize these using a neural network together with a manifold-constrained variable.

The code space \(\mathcal{C} \subset \mathbb{C}^{2^n}\) has dimension $2^k$ and is represented by an isometry
\begin{equation}
U_\mathcal{C} \in \mathbb{C}^{2^n \times 2^k}, \qquad U_\mathcal{C}^\dagger U_\mathcal{C} = I_{2^k},
\end{equation}
whose columns form an orthonormal basis spanning $\mathcal{C}$. The matrix $U_\mathcal{C}$ is treated as a trainable parameter constrained to lie on the Grassmannian manifold \[\mathcal{M} = \mathrm{Gr}(\mathbb{C}^{2^n}, \mathbb{C}^{2^k}).\]

\begin{figure}[H]
    \centering
    \includegraphics[width=0.9\linewidth]{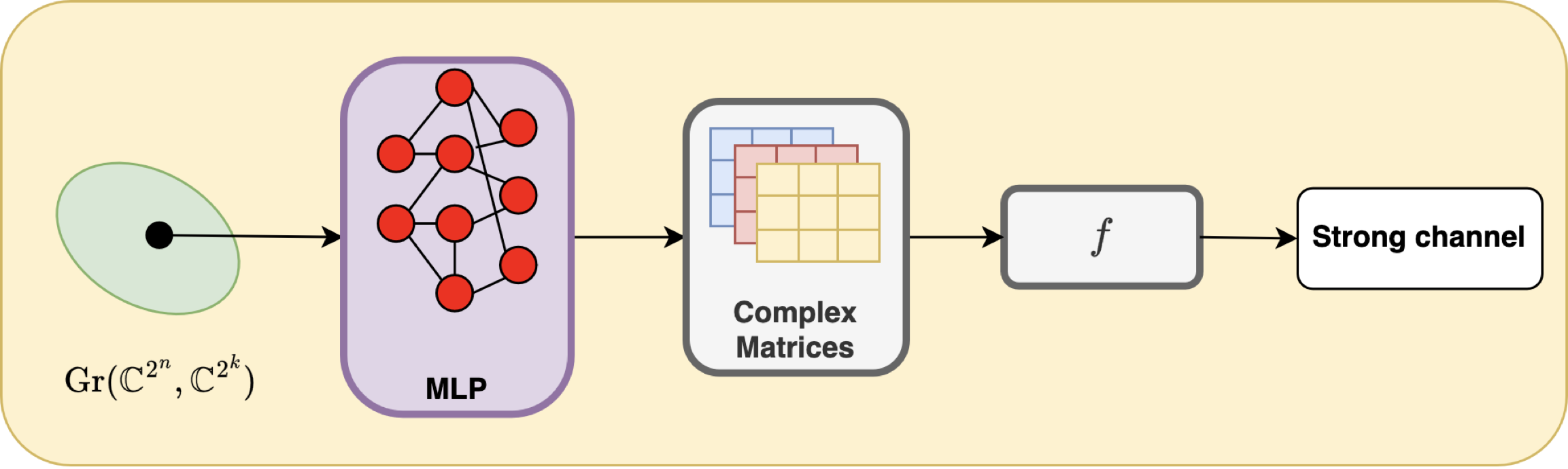}
    \caption{In the forward pass, the code space (represented by a point in $\mathcal{M}$) is input to an MLP, outputting a set of $L$ complex matrices. A nonlinear activation function $f$ then returns a set of Kraus operators for the associated recovery channel.}
    \label{fig:nn_arch}
\end{figure}

We define a multilayer perceptron (MLP), parameterized by $\Theta$, that maps the code space representation $U_\mathcal{C}$ to a set of Kraus operators describing a quantum channel:
\[
    f_\Theta: \mathcal{M} \rightarrow \mathbb{C}^{L \times 2^n \times 2^n}.
\]
The network applies six fully connected hidden layers, each of width 128 with Rectified Linear Unit (ReLU) activations. The final linear layer outputs $L \times (2^n \times 2^n)$ parameters, which are reshaped into $L$ complex matrices
\begin{equation}
\textbf{G} = \{G_\ell\}_{\ell=1}^L, \qquad G_\ell \in \mathbb{C}^{2^n \times 2^n}.
\end{equation}
These matrices are mapped through a nonlinear function $f$ (defined later) to produce a valid set of Kraus operators
\begin{equation}
\mathbf{K} = f(\mathbf{G}) = \{K_\ell\}_{\ell=1}^L,
\end{equation}
which define a quantum channel, i.e., a completely-positive and trace-preserving (CPTP) map. Since the set of CPTP maps is convex \cite{verstraete_quantum_2003, chou_five_2022, kukulski_generating_2021}, the channels can be parameterized by a set of complex matrices. Therefore, every step of backward propagation yields a different quantum channel. In contrast, the set of code spaces is not convex. Accordingly, optimization for $U_\mathcal{C}$ is performed using Riemannian gradient descent \cite{bonnabel_stochastic_2013, becigneul_riemannian_2019, huang_building_2018, smith_optimization_2014}, which preserves the orthonormality constraint throughout training. The neural network architecture is illustrated in \cref{fig:nn_arch}.

\subsection{Characterizing weak channels}\label{sec:weak_channel_char}

A strong channel can be used to find a family of weak channels parameterized by a small number \[\epsilon=\sqrt{\kappa\delta t}\] simply by taking a convex combination of it with the identity channel:
\begin{equation}\label{eq:convert_strong_to_weak_jump}
    \tilde{\mathcal{R}} = (1-\epsilon^2)\mathcal{I} + \epsilon^2\mathcal{R}.
\end{equation}
Such a channel employs Kraus operators
\[
\begin{split}
    \tilde{R}_0 &= \sqrt{1-\epsilon^2}I, \\
    \tilde{R}_\ell &= \epsilon R_\ell,
\end{split}
\]
where $\{R_\ell\}_{\ell=1}^{L}$ is a Kraus decomposition of $\mathcal{R}$. The problem of characterizing the weak channel $\tilde{\mathcal{R}}$ reduces to that of a strong channel $\mathcal{R}$. Consider a set of arbitrary complex matrices $\textbf{G} = \{G_\ell\}_{\ell=1}^{L}$ of dimension $2^n \times 2^n$. Compute the positive semi-definite matrix \[H = \sum_\ell G_\ell^\dagger G_\ell \geq 0.\] The matrices \[R_\ell=G_\ell H^{-1/2}\] form a valid set of Kraus operators for a quantum channel \cite{kukulski_generating_2021}. Observe that 
\[
\sum_\ell R_\ell^\dagger R_\ell = H^{-1/2}HH^{-1/2} = I_{2^n} ,
\]
showing that the CP map is trace-preserving. The arbitrary operators $\textbf{G}$ are obtained from the final layer of a neural network. We train this neural network to yield a locally optimal weak recovery channel $\tilde{\mathcal{R}}$.

The master equation describing the state evolution with the obtained weak channel is
\begin{equation}\label{eq:unconditional_master}
    \frac{\mathrm{d}\rho}{\mathrm{d}t} = \kappa [\mathcal{R}(\rho) - \rho].
\end{equation}
To convert this into a more practical protocol, we can decompose the Kraus map into a weak measurement followed by a weak correction, yielding continuous measurement with feedback in the continuous-time limit. For example, one particular decomposition is
\begin{equation}\label{eq:kraus_diff}
    \tilde{K}_{\ell, \pm} = \frac{1}{\sqrt{2L}}\left(I \pm \epsilon' R_\ell - \frac{\epsilon'^2}{2}R_\ell^\dagger R_\ell\right) = \tilde{U}_{\ell, \pm}\tilde{M}_{\ell, \pm},
\end{equation}
where $\epsilon'=\epsilon\sqrt{L}$, and the second equality is the polar decomposition. Observe that $\{\tilde{M}_{\ell, \pm}\}$ forms a weak POVM, and $\{\tilde{U}_{\ell, \pm}\}$ are the associated weak unitary corrections. The two
decompositions $\{\tilde{R}_\ell\}$ and $\{\tilde{K}_{\ell, \pm}\}$ correspond
to the same weak channel, up to the order relevant in the continuous-time
limit, and hence they perform the same \emph{on average}. However, in a single
unraveling, the former is a jump-like weak channel that usually leaves the
state unaffected, but with a small probability makes the state jump. The latter
is a diffusive-type weak channel that stochastically perturbs the state by a
small amount at each step.

This diffusive unraveling has an operational interpretation as a homodyne-type continuous measurement. In a homodyne measurement, a single quadrature of the output field is monitored continuously, producing a real-valued noisy measurement current. Such measurements are native to superconducting transmon-based circuit-QED architectures: transmon information is dispersively mapped onto microwave resonator quadratures, and the outgoing microwave field is then amplified and measured by homodyne or heterodyne detection \cite{murch2013observing,blais2021circuit,hacohen2020continuous}. Thus the diffusive recovery map can be implemented as weak quadrature measurements
followed by Markovian, current-dependent feedback of the Wiseman--Milburn type
\cite{wiseman1993feedback,wiseman2010quantum}.

We decompose each recovery Kraus operator into its Hermitian and anti-Hermitian
parts as
\[
    R_\ell = O_\ell - iH_\ell,
\]
where
\[
    O_\ell = \frac{R_\ell+R_\ell^\dagger}{2}, \qquad
    H_\ell = i\frac{R_\ell-R_\ell^\dagger}{2}.
\]
Here $O_\ell=O_\ell^\dagger$ is the measured quadrature, and
$H_\ell=H_\ell^\dagger$ is the outcome-dependent correction. In the continuous-time limit, the measurement current corresponding to the observable $O_\ell$ is
\[\label{eq:measurement_current}
    J_\ell(t) = \langle O_\ell\rangle_c + \frac{1}{2\sqrt{\kappa}}\frac{\mathrm{d}W_t}{\mathrm{d}t}.
\]

The Hamiltonian control applied in response to the homodyne current is
\begin{equation}\label{eq:total_feedback_hamiltonian}
    H_{\mathrm{ctrl}}(t)
    =
    H_{\mathrm{det}}
    +
    H_{\mathrm{fb}}(t),
\end{equation}
where the deterministic part is
\begin{equation}
    H_{\mathrm{det}} = \sum_{\ell=1}^{L} H_\ell^{(0)}, \quad H_\ell^{(0)} = -\frac{\kappa}{2} \{H_\ell,O_\ell\},
\end{equation}
and the Markovian current-dependent feedback Hamiltonian is
\begin{equation}\label{eq:feedback_hamiltonian}
    H_{\mathrm{fb}}(t)
    =
    2\kappa
    \sum_{\ell=1}^{L}
    J_\ell(t)H_\ell.
\end{equation}
Equivalently, one may absorb $H_{\mathrm{det}}$ into the background system
Hamiltonian and regard \cref{eq:feedback_hamiltonian} as the real-time feedback
term.

For ideal unit-efficiency monitoring, the resulting conditional master equation
in It\^o form can be written in terms of the measured quadratures $O_\ell$ and feedback generators $H_\ell$ as
\begin{equation}\label{eq:conditional_master}
\begin{split}
    \mathrm{d}\rho_c 
    &= \sum_{\ell=1}^{L} \bigg[ -i[H_\ell^{(0)},\rho_c]\,\mathrm{d}t - i\kappa[H_\ell,\{O_\ell,\rho_c\}]\,\mathrm{d}t \\
    &\qquad + \kappa\mathcal{D}[O_\ell]\rho_c\,\mathrm{d}t + \kappa\mathcal{D}[H_\ell]\rho_c\,\mathrm{d}t \\
    &\qquad + \sqrt{\kappa}\mathcal{H}[O_\ell]\rho_c\,\mathrm{d}W_t^{(\ell)} - i\sqrt{\kappa}[H_\ell,\rho_c]\,\mathrm{d}W_t^{(\ell)} \bigg],
\end{split}
\end{equation}
The derivation is given in \cref{app:conditional_master}. Taking the ensemble average of \cref{eq:conditional_master} over measurement trajectories recovers the unconditional recovery dynamics \cref{eq:unconditional_master}.

\subsection{Characterizing the code space}\label{sec:cs_char}

For a QECC with $n$ physical qubits and $k$ logical qubits, the code space is a subspace of dimension $2^k$ in the ambient space $\mathbb{C}^{2^n}$. Searching for a ``good'' code space can be done over the set of all $2^k$-dimensional subspaces. This set is not convex, and forms the Grassmannian manifold $\mathcal{M}$ with metric $g$. A point $x \in \mathcal{M}$ can be represented by a $2^n \times 2^k$ complex-valued matrix $U_x$ with orthonormal columns forming a basis for $x$. The tangent space of $\mathcal{M}$ at $x$, $\mathcal{T}_x^{\mathcal{M}}$ consists of matrices $Z \in \mathbb{C}^{2^n \times 2^k}$ satisfying $U_x^\dagger Z = 0$, with orthogonal projector \[\Pi_{\mathcal{T}_x^{\mathcal{M}}} = I - U_x U_x^\dagger.\] The Riemannian gradient of a smooth function $\mathcal{L}: \mathcal{M} \rightarrow \mathbb{R}$ at $x$ is obtained by projecting its Euclidean gradient onto $\mathcal{T}_x^\mathcal{M}$:
\begin{equation}
    \nabla_x^{\mathcal{M}}f = \Pi_{\mathcal{T}_x^\mathcal{M}} \nabla \mathcal{L}.
\end{equation}
To perform gradient descent, the point in $\mathcal{M}$ is moved in the direction of the negative gradient along the geodesic. However, the \emph{exponential map} $\exp_x(v)$ to move along the geodesics occasionally induces numerical errors that move the point out of the manifold; one has to project it back using the polar decomposition. This procedure is computationally expensive and not really necessary, since the step size is small. A good approximation is to move along a straight line and then \emph{retract} back on to the manifold. One step of steepest descent by this approach would be
\begin{equation}
    U_x := \text{Ret}\left(U_x - \eta \nabla_x^\mathcal{M}f\right) = \tilde{U}\tilde{V}^\dagger,
\end{equation}
where $\tilde{U}$ ($\tilde{V}$) are the left (right) singular vectors and $\eta$ is the learning rate.

\subsection{Cost function}\label{sec:cost}

The objective is to maintain the \emph{average logical state} as long as possible. Consider the maximally entangled state between a logical qubit in the code $\mathcal{C}$ (denoted by a bar) and a \emph{static reference} qubit:
\begin{equation}\label{eq:phi_cr}
\ket{\Phi}_{LR}^{\mathcal{C}} = \sum_{j=0}^{2^k-1} \frac{\ket{\bar{j}} \otimes  \ket{j}}{\sqrt{2^k}},
\end{equation}
where $j$ (inside the ket) is in the binary notation. Tracing out the reference qubit gives an equal mixture of all logical states. Consider the dynamics induced on $\rho_{LR}(0) = \ket{\Phi}^{\mathcal{C}}_{LR}\bra{\Phi}^\mathcal{C}$, subject to a weak Markovian noise superoperator $\mathcal{L}_n$ and a weak recovery channel $\tilde{\mathcal{R}}$ on $n$ qubits in a small time-interval $\delta t$:
\begin{eqnarray}
\rho_{LR}(\delta t) &=& (1-\kappa\delta t)\rho_{LR}(0) + (\kappa\delta t)\mathcal{R}\left[\rho_{LR}(0)\right] \nonumber\\
&& + \mathcal{L}_n\left[\rho_{LR}(0)\right] \delta t .
\end{eqnarray}
Neither the noise nor the recovery act on the reference system. Since $\rho_{LR}(0) \in \mathcal{C}$, applying $\mathcal{R}$ in the second term leaves it invariant. This yields
\begin{equation}
\rho_{LR}(\delta t) = \rho_{LR}(0) + \mathcal{L}_n\left[\rho_{LR}(0)\right] \delta t .
\end{equation}
For QEC to be beneficial, we would like the code $\mathcal{C}$ to preserve its $k$ logical qubits at least as long as $k$ physical qubits would last under an analogous noise $\mathcal{L}_k$. However, since $n>k$, the noise $\mathcal{L}_n$ has a stronger effect than $\mathcal{L}_k$, producing a transient period during which the fidelity drops faster for the encoded qubits than the ``bare'' qubits. This is analytically confirmed in \cite{ippoliti_perturbative_2015}. For the case of optimizing the recovery map, choosing the objective function as the infidelity between $\ket{\Phi}_{LR}$ and $\rho_{LR}(\delta t)$ yields the trivial identity map, as expected by this argument. But this fails to protect the logical states over longer times; most of the drop in fidelity for the encoded qubits could be correctable errors, while all errors on ``bare'' qubits are uncorrectable. To avoid this problem, we instead define the cost function by applying the strong recovery map to $\rho_{LR}(\delta t)$ before calculating the infidelity. This gives
\begin{equation}\label{eq:strong_rec}
\mathcal{R}\left[\rho_{LR}(\delta t)\right] = \rho_{LR}(0) + \mathcal{R}\left[\mathcal{L}_n\left[\rho_{LR}(0)\right]\right] \delta t,
\end{equation}
where the strong recovery in the second term mitigates the drop in fidelity due to correctable errors, slowing the decay of the encoded $n$-qubit state compared to the bare $k$-qubit state under $\mathcal{L}_k$. Hence, we choose to minimize the infidelity between \cref{eq:phi_cr} and \cref{eq:strong_rec}. To this end, we choose the objective function for Markovian noise environments as
\begin{equation}\label{eq:cost}
    \mathcal{L}^\mathrm{M}(\mathcal{C}, \mathcal{R}) = 1 - \bra{\Phi}_{LR}^{\mathcal{C}} \mathcal{R}\left[\rho_{LR}(\delta t)\right] \ket{\Phi}_{LR}^{\mathcal{C}}.
\end{equation}

When the noise is correlated in time (non-Markovian), however, this objective function is not directly applicable. A natural alternative is to unravel the evolution into a collection of trajectories and minimize the sum of infidelities over the ensemble average through time:
\begin{equation}
\mathcal{L}^{\mathrm{NM}}(\mathcal{C}, \mathcal{R}) = 1 - \frac{1}{T}\int_0^T \bra{\Phi}_{LR}^{\mathcal{C}} \bar{\rho}_{LR}(t) \ket{\Phi}_{LR}^{\mathcal{C}}\, \mathrm{d}t,
\end{equation}
where $\bar{\rho}_{LR}(t)$ is the ensemble average over the noise and recovery.

\section{Examples}\label{sec:examples}

We considered several noise processes $\mathcal{L}_n$ and compared the average code state fidelity between 1) optimized code space and recovery 2) standard code space and recovery 3) bare $k$ qubits with no recovery. These examples were chosen to include noise that goes beyond independent, unital, Markovian noise on the qubits.

A standard strong recovery map has a set of Kraus operators $\{U_sP_s\}_{s=1}^{2^{n-k}}$, where $s$ labels the error syndromes, $P_s$ is the projector onto a syndrome space and $U_s$ is the unitary to rotate from the corresponding syndrome space to the code space. This strong map can be converted into a weak map using \cref{eq:convert_strong_to_weak_jump}. Training is performed with $3$ different seeds.

\subsection{Markovian decoherence}

\subsubsection{Bit-flip}

Three qubits subject to uniform Markovian bit-flip noise, whose dynamics can be modeled using a Lindblad master equation with the Lindblad operators $\{\sqrt{\gamma}X_i\}$, where $X_i$ is the Pauli $X$ operator acting on qubit $i$. Here and hereafter, $\gamma$ is the error rate. Remarkably, the code discovered by the neural network performs as well as the standard $[\![3, 1, 3]\!]$ bit-flip code, as shown in \cref{fig:infidelity_bitflip_3q}. This gives empirical evidence that the $[\![3, 1, 3]\!]$ code is optimal for protecting against such symmetrical noise. However, as we shall see, the neural network learns to find better-performing codes and recoveries for more complex noise models.

\subsubsection{Amplitude damping and correlated dephasing}


We next consider five physical qubits encoding one logical qubit and subject to both Markovian amplitude damping and \emph{spatially} correlated dephasing. Specifically, the Lindblad operators are $\sqrt{\gamma}\sigma_j^-, j=1,\ldots,5,$ together with $\sqrt{\gamma}Z_1Z_2, \sqrt{\gamma}Z_3Z_4, \sqrt{\gamma}Z_5$, where, with the convention used here, $\sigma_j^- = X_j-iY_j$. We compare our jointly optimized code--recovery pair with the perfect $[\![5,1,3]\!]$ code, which corrects arbitrary single-qubit Pauli errors under its standard recovery.

This example also provides a comparison of the two components of our optimization. In particular, we compare three strategies: (i) the standard $[\![5,1,3]\!]$ code with its standard single-qubit Pauli recovery; (ii) the same fixed code with a modified recovery for the present noise process; and (iii) the code space and recovery channel optimized jointly by our method. The first two strategies use exactly the same code space, and therefore isolate the improvement that can be obtained by adapting the recovery while holding the code fixed. The comparison with the third strategy then shows the additional benefit obtained by adapting the code space and recovery together.

For the standard $[\![5,1,3]\!]$ code, the errors $Y_1$ and $Z_3Z_4$ produce the same syndrome, as do $Y_4$ and $Z_1Z_2$. In each case, the two errors are inequivalent modulo the stabilizer and differ by a nontrivial logical operation on the code space. Consequently, if $P$ denotes the projector onto the standard code space and $E_a$ and $E_b$ denote either of these
pairs, then
\[
    P E_a^\dagger E_b P \not\propto P.
\]
The Knill--Laflamme conditions therefore preclude the simultaneous perfect correction of both errors by any recovery channel acting on this fixed code. A channel-adapted recovery can choose which member of each syndrome-colliding pair to favor, according to its relative importance in the noise process, but it cannot remove this intrinsic ambiguity. This is particularly relevant here because amplitude damping contains both $X$ and $Y$ components, so favoring the correlated dephasing errors comes at the cost of reduced correction of some amplitude-damping errors.

As shown in \cref{fig:infidelity_correlated_5q}, the standard $[\![5,1,3]\!]$ recovery performs the worst. Adapting the recovery to correct the dephasing errors for the colliding syndromes while holding the $[\![5,1,3]\!]$ code fixed produces a substantial improvement. However, the jointly optimized code-recovery pair achieves still lower logical infidelity. This suggests that joint optimization can change the geometry of the code space so that the relevant noise-induced subspaces are more distinguishable, while simultaneously adapting the recovery to those new subspaces. This example shows that optimizing the recovery alone for a fixed code need not attain the best achievable performance.

\subsubsection{Leakage}

We consider leakage errors out of the computational subspace. We model this situation by encoding one logical qubit using two physical qubits. Each physical qubit also has an additional basis state, $\ket{\mathrm{leak}}$, to which erroneous transitions can occur. The Lindblad operators are $\sqrt{\gamma} \ket{\mathrm{leak}}_j \bra{1}$, where the index $j=1,2$ denotes each physical qubit. The code space thus forms a two-dimensional subspace within the four-dimensional subspace spanned by $\{\ket{00},\ket{01},\ket{10},\ket{11}\}$ within $\mathbb{C}^{3^2}$. The recovery Kraus operators can include the leakage state, since it may be utilized to perform measurements. \cref{fig:infidelity_leakage_2q} shows that the ML approach can mitigate leakage errors even in this low-dimensional system, and greatly outperforms an unencoded physical qubit.

\subsubsection{Qutrit CT-QEC}

\begin{figure*}[t]
    \centering

    \subfigure[]{%
        \includegraphics[width=0.3\textwidth]{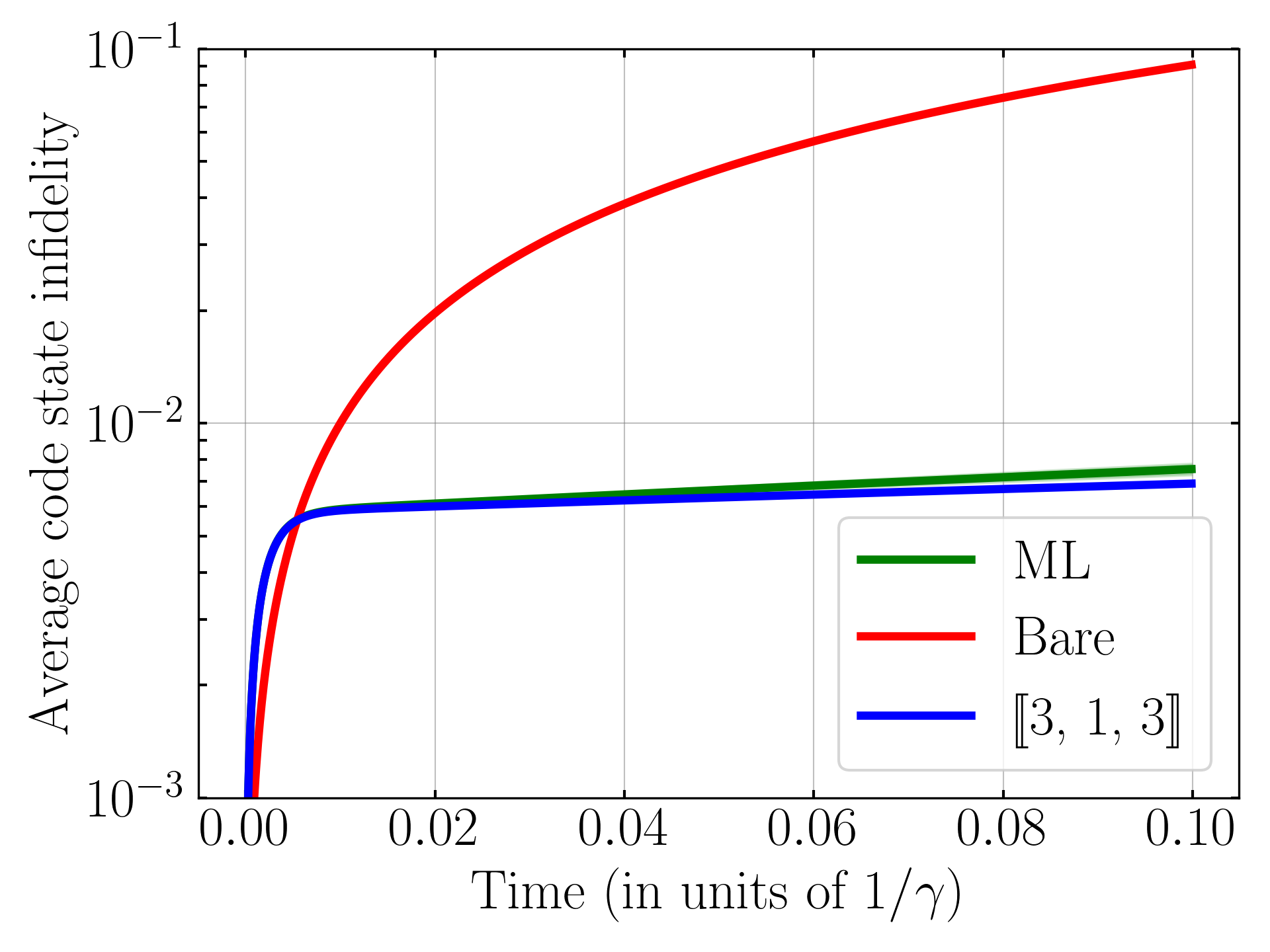}
        \label{fig:infidelity_bitflip_3q}
    }
    \subfigure[]{%
        \includegraphics[width=0.3\textwidth]{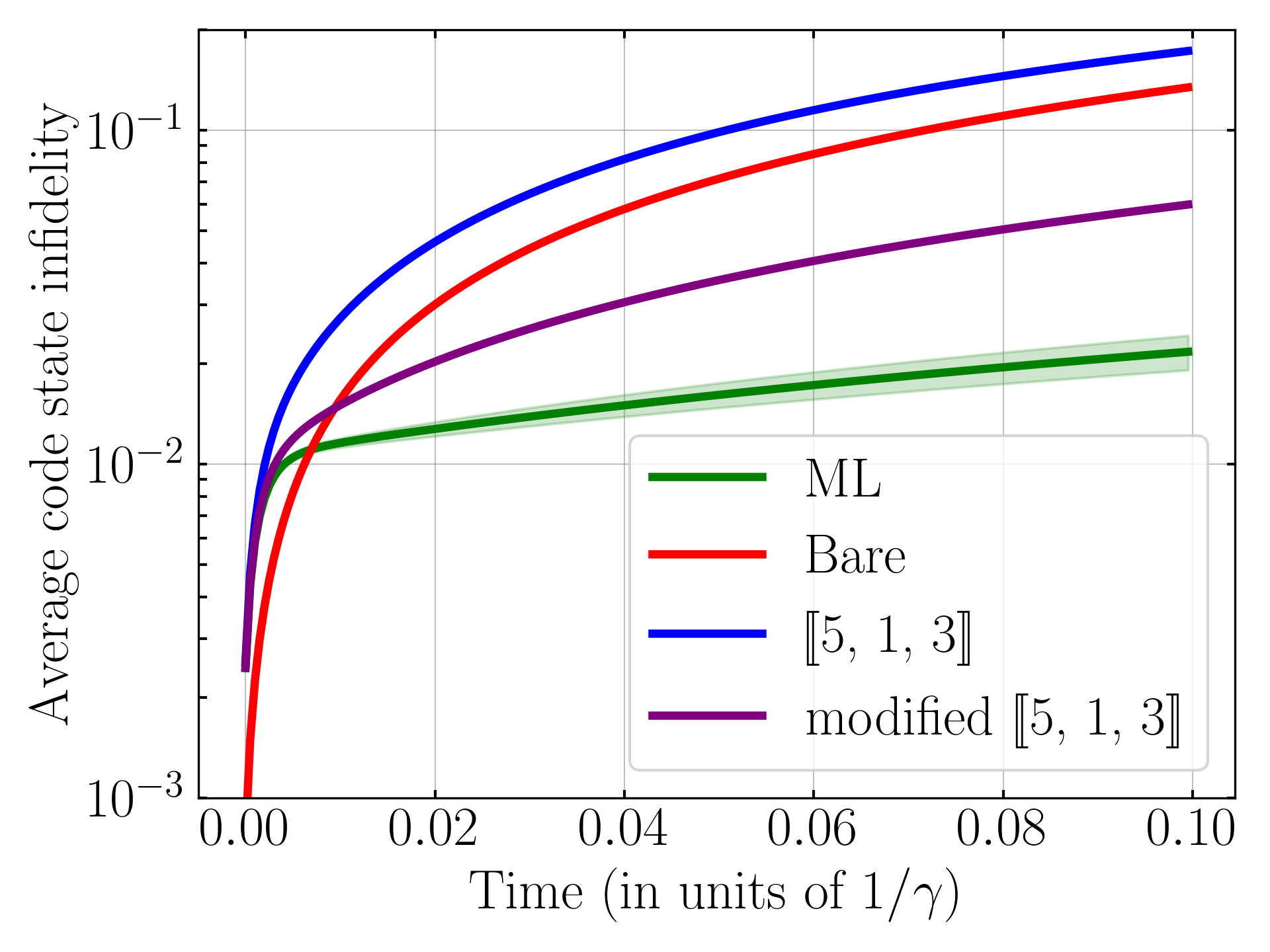}
        \label{fig:infidelity_correlated_5q}
    }
    \subfigure[]{%
        \includegraphics[width=0.3\textwidth]{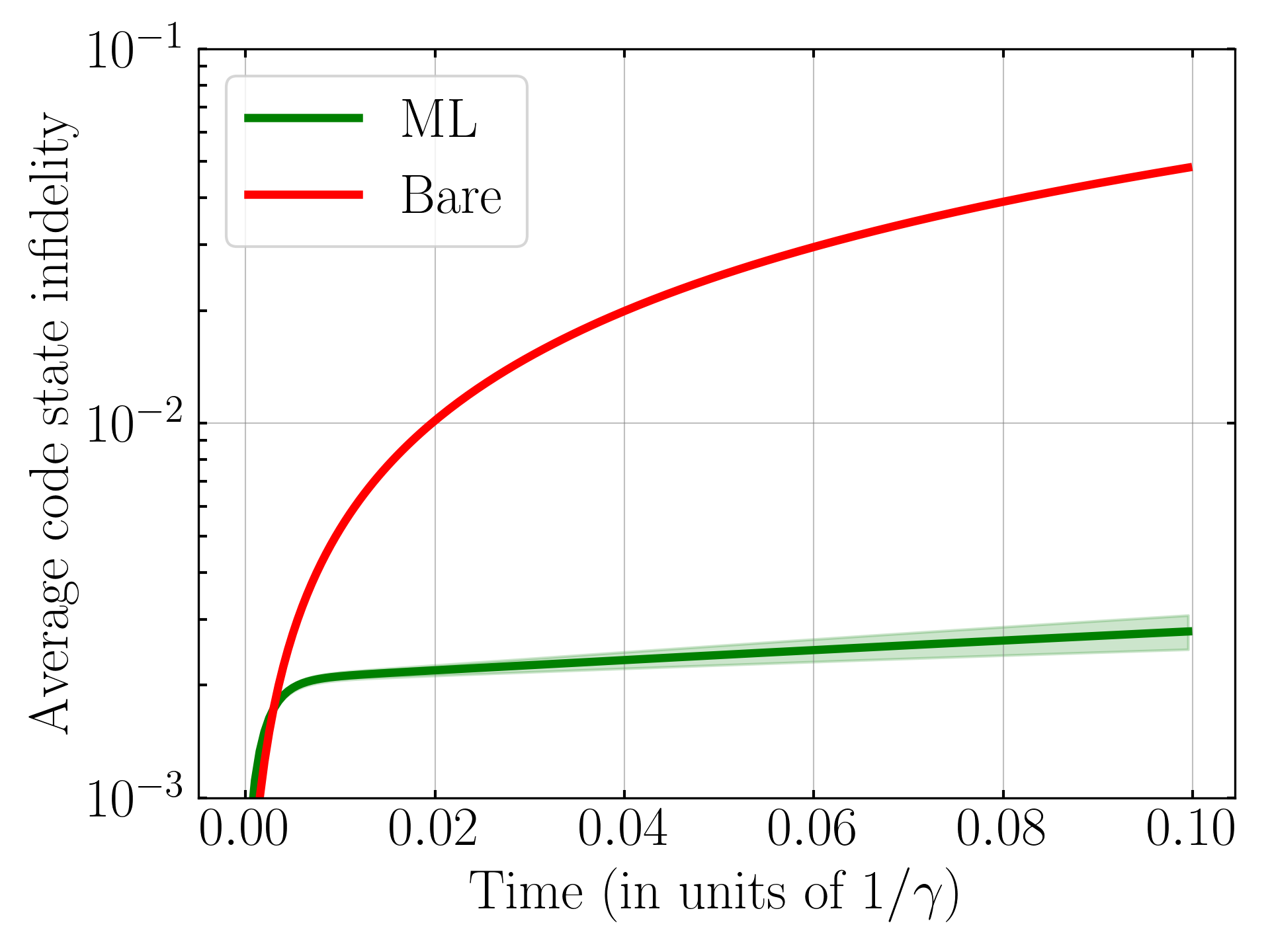}
        \label{fig:infidelity_leakage_2q}
    }
    
    \subfigure[]{%
        \includegraphics[width=0.3\textwidth]{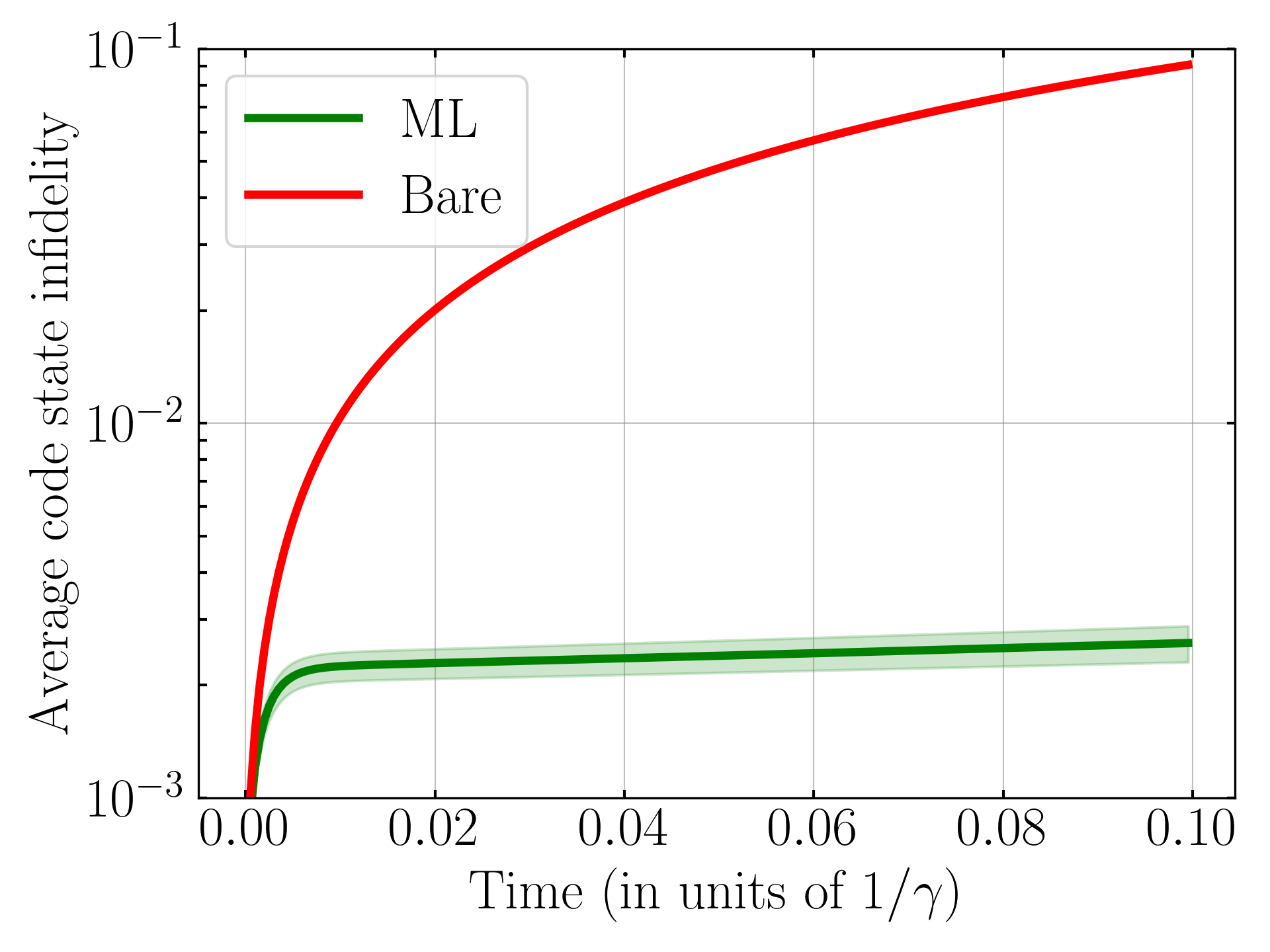}
        \label{fig:infidelity_qutrit_2q}
    }
    \subfigure[]{%
        \includegraphics[width=0.3\textwidth]{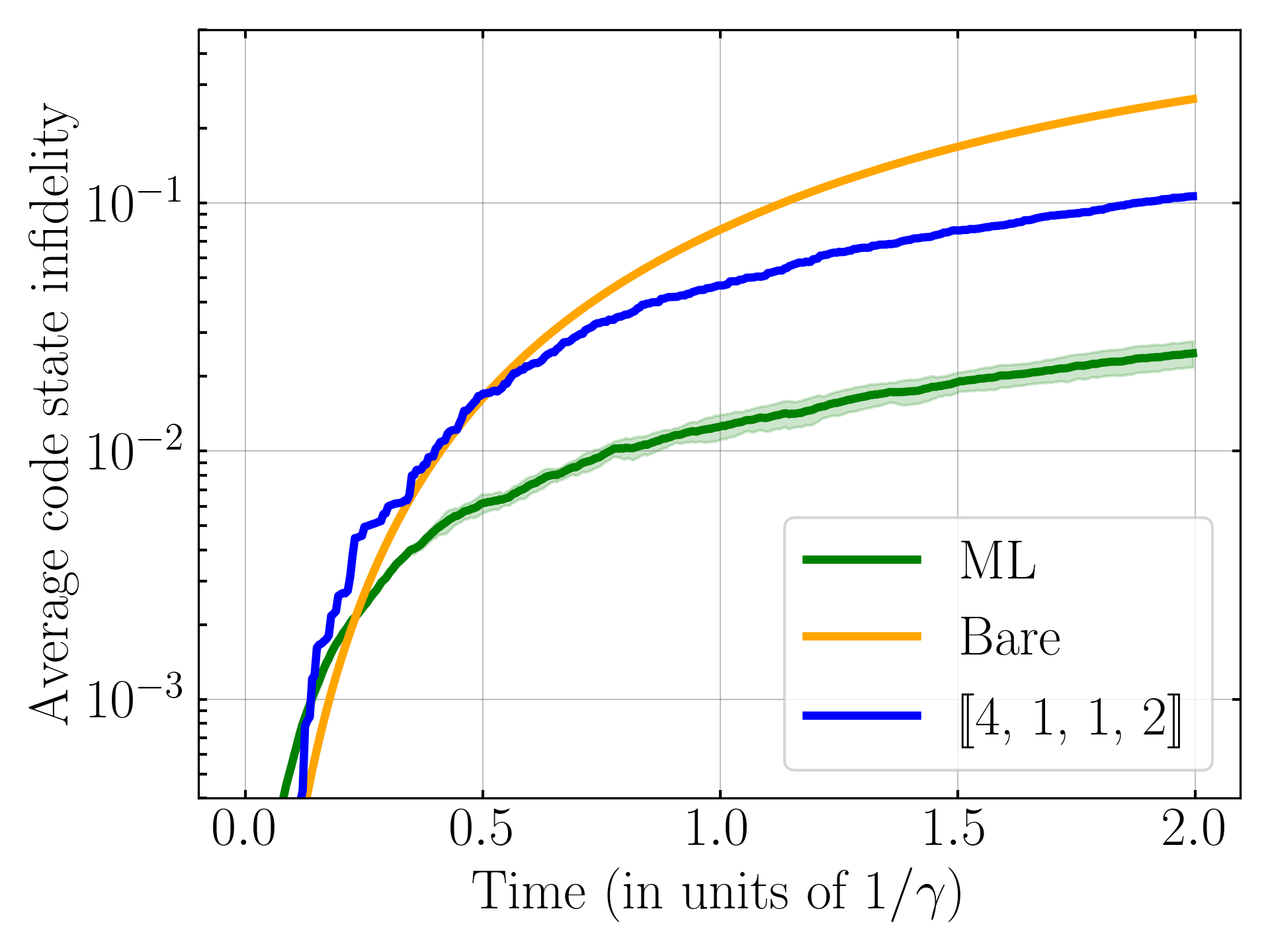}
        \label{fig:infidelity_low_freq_4q}
    }
    \subfigure[]{%
        \includegraphics[width=0.3\textwidth]{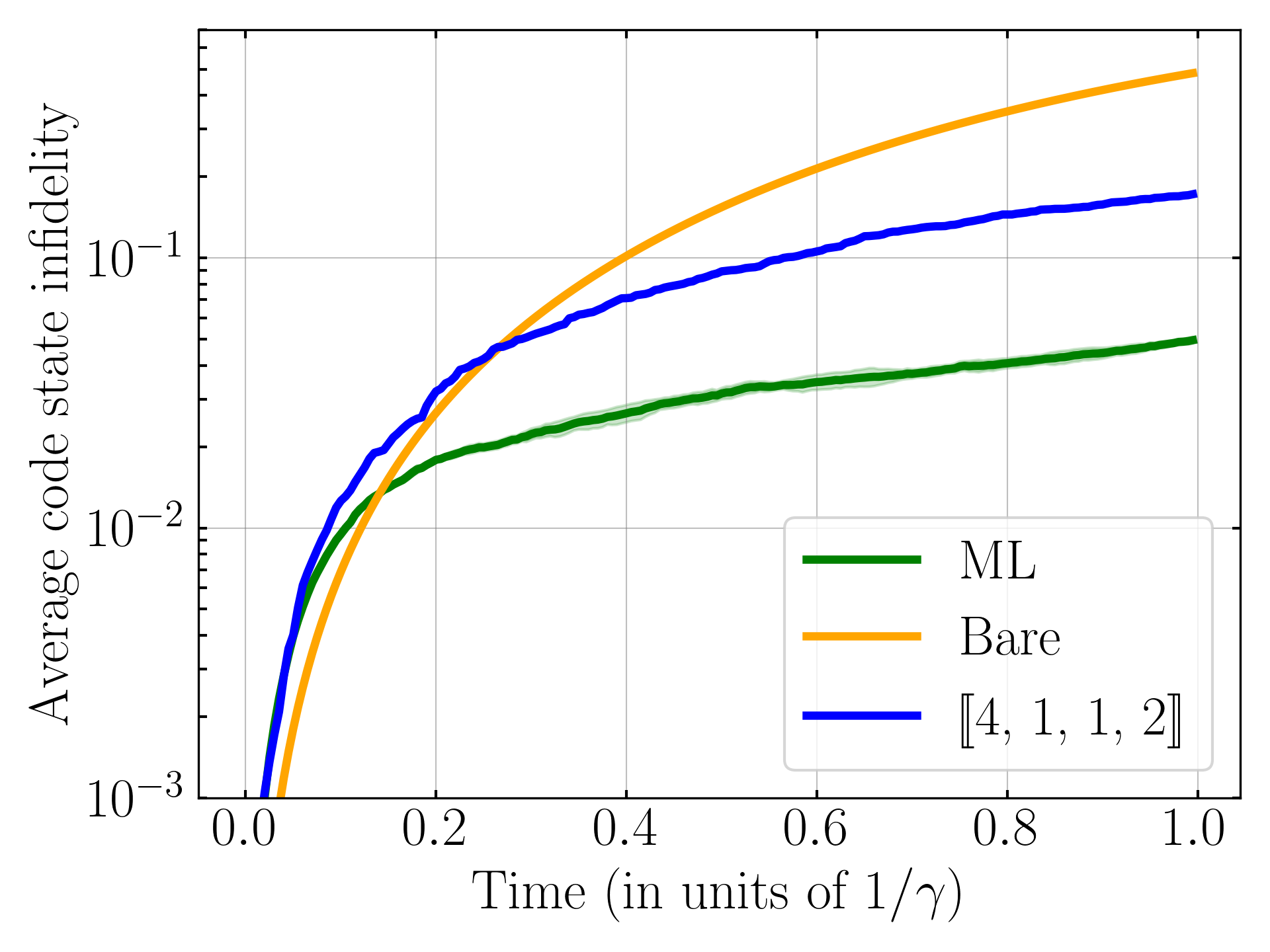}
        \label{fig:infidelity_constant_4q}
    }

    \caption{Average code state fidelity decay as a function of time in units of $1/\gamma$. Noise process is (a) Markovian bit-flip noise on 3 qubits; (b) Amplitude damping and spatially correlated dephasing on 5 qubits; the difference between the two $[\![5,1,3]\!]$ curves shows the gain from adapting the recovery while holding the code fixed, whereas the remaining gap to the jointly optimized result shows the additional gain from co-adapting the code space and recovery channel. (c) leakage out of computational subspace in 2 qubits; (d) population transitions in lowest adjacent levels of 2 qutrits; (e) non-Markovian $1/f$ noise of type $X$ and $Z$ on 4 qubits; and (f) constant Hamiltonian errors of type $X$ and $Z$ on 4 qubits. Standard deviation is from 3 initial training seeds.}
    \label{fig:infidelities}
\end{figure*}

The ML method makes it straightforward to encode logical qubits in a qutrit-based system. Specifically, we encode a single logical qubit using two physical qutrits. We model the noise as population transitions between the lowest energy levels; the associated Lindblad operators are $\sqrt{\gamma}\{\ket{0}_j\bra{1} + \ket{1}_j\bra{0}\}$, where the index $j=1,2$ denotes the qutrit. Since this code provides a logical qubit, we compare its performance to a bare physical qubit subject to $\sqrt{\gamma}X$ noise. The average code state infidelity is shown in \cref{fig:infidelity_qutrit_2q}.

\subsection{Non-Markovian decoherence}

\subsubsection{$1/f$ noise}

The errors are continuous-in-time $1/f$ Hamiltonian rotations:
\begin{equation}\label{eq:low_freq_ham}
H_{\mathrm{err}}(t) = \sum_{j} \lambda_{j,X}(t) X_j + \lambda_{j,Z}(t)Z_j,
\end{equation}
where $j$ labels each qubit and $\lambda_{j,\alpha}(t)$ (for $\alpha=X,Z$) is a time-dependent scalar function. Each $\lambda_{j,\alpha}$ is a sum of exponentially-decaying random pulses occuring at random times with a constant rate $\gamma$:
\begin{equation}
\lambda_{j,\alpha}(t) = \sum_{K} \epsilon_K \theta(t-t_{j,\alpha}) \exp\left(-\frac{t-t_{j,\alpha}}{\tau}\right).
\end{equation}
Here, the index $K$ labels the pulses, $\epsilon_K \sim U(-\epsilon, \epsilon)$ where $\epsilon$ is the maximum pulse amplitude and $U$ denotes the uniform distribution, $\theta(t)$ is the Heaviside step function and $\tau$ is the pulse lifetime. We choose the inter-arrival times in $\lambda_{j\alpha}(t)$ to be independent and exponentially distributed (i.e., a Poisson process), $\Delta t_{j,\alpha} \sim \exp(t;\gamma)$.

We simulated an example where 1 logical qubit is encoded in 4 physical qubits, and the error Hamiltonian contains both Pauli $X$ and $Z$. We compared the performance to both a bare qubit and the 4-qubit Bacon-Shor code. While it is only an error-detecting code, continuously measuring the stabilizers should suppress the non-Markovian noise due to the quantum Zeno effect \cite{chen_continuous_2020}. The neural network, however, found a significantly better code and recovery tailored for this noise process, as shown in \cref{fig:infidelity_low_freq_4q}.

\subsubsection{Constant Hamiltonian errors}

The stochastic Hamiltonian is modeled as \[H_\mathrm{err} = \sum_{j, \alpha} \lambda \sigma_j^\alpha.\] This is a simple model that is even more non-Markovian than the $1/f$ noise model. Again, we used $4$ physical qubits to encode $1$ logical qubit against Pauli-$X$ and $Z$ rotations. Since this is similar to the previous noise model, we reused the same code and recovery to test against this model. Notably, it performs better than the $[\![4, 1, 1, 2]\!]$ code (and the bare qubit), as shown in \cref{fig:infidelity_constant_4q}, in spite of not being trained specifically for this noise process. This suggests that the ML method is not overfitting to the noise parameters but can generalize well to similar noise models. Furthermore, we observe that the standard deviations across seeds are low, indicating that our method is robust to random initial parameters. 

\section{Discussion}\label{sec:discussion}

\textit{Summary---}We presented a neural-network-based method to identify optimal quantum error correcting codes and their associated weak recovery channels within the CT-QEC framework. Conventional quantum error correction typically use stabilizer codes due to their ease of description in terms of Pauli operators. Because any correctable error can be expressed as a linear combination of such operators, stabilizer codes can, in principle, accommodate arbitrary noise models. However, codes tailored to the structure of specific, nontrivial noise processes can outperform stabilizer constructions. This is particularly apparent in the Noisy Intermediate-Scale Quantum (NISQ) regime, where noise is often highly asymmetric and device dependent. Our protocol systematically identifies codes and recovery operations that outperform stabilizer codes under such conditions.

We show that the code space and recovery channel should be treated as a matched pair. Optimizing only the recovery restricts the calculation to the error-distinguishability properties of a fixed code. If that code maps two relevant errors to the same corrupted subspace with different logical actions, no subsequent recovery can perfectly distinguish and correct both. Conversely, optimizing the code while holding the recovery fixed evaluates candidate code spaces through a decoder whose measurement and correction operators were designed for a different set of noise induced subspaces. Such a fixed recovery can therefore make a favorable code appear unfavorable. Joint optimization avoids both restrictions by adapting the code space and recovery to one another according to the final logical performance objective.

We evaluated the approach across a diverse set of noise models, including uniform Markovian bit-flip errors; amplitude damping and correlated dephasing; leakage from the computational subspace; generalized Pauli errors on qutrits; and non-Markovian noise sources such as static and pink ($1/f$) noise. Although we demonstrated the method on relatively simple examples, the framework extends naturally to more realistic noise channels. In particular, it supports a data-driven workflow in which codes, measurement observables, and feedback Hamiltonians are optimized directly from experimentally characterized device noise.

\textit{Future directions---}The current framework can serve as a tool to probe theoretical limits of error-correcting performance for a given noise process. In practice, the recovery maps obtained from the neural network can be realized via continuous measurements combined with real-time feedback Hamiltonians. In the present work, we did not impose constraints on the geometry of the recovery channels; as a result, the corresponding measurements and Hamiltonians may be high-weight, nonlocal, or otherwise difficult to implement. In principle, device-level constraints---such as locality or coupling structure---could be incorporated into the loss function. While such constraints may reduce achievable performance, they would make the protocol more practical. Exploring this trade-off remains an important direction for future work.

The size of the neural networks grows exponentially with system size, rendering training increasingly impractical for larger systems. Consequently, the present approach is not directly scalable. Its primary utility lies in identifying codes tailored to device-specific noise. Because the protocol operates in continuous time, temporally correlated errors are effectively suppressed, so that the dominant residual errors are Markovian. Standard error-correcting codes may then be concatenated with these optimized encodings to achieve a desired level of error correction.

The optimization of discrete-time quantum error correction has been explored using convex and biconvex techniques \cite{kosut_quantum_2009, wang_automated_2022, zeng_approximate_2023}, and optimality conditions for the well-known Petz recovery map were established in \cite{petz_optim_cond}. In the continuous-time setting, \cite{petz_lindblad} implements quantum error correction via the Petz map for Markovian noise processes. It is plausible that related optimization frameworks could also be used to construct weak recovery maps of the type considered here. A systematic comparison between these approaches and the present method remains for future work.

\acknowledgements

We thank Prithviraj Prabhu, Juan Garcia-Nila, and Daniel Lidar for useful discussions. This work was supported in part by the U. S. Army Research Laboratory and the U. S. Army Research Office under contract/grant number W911NF2310255, and by NSF Grant FET-2316713.

\section*{Data Availability}

The code used to train and evaluate the models used in this paper is available at \cite{code}.


\bibliography{references}

\appendix

\begin{widetext}

\section{Continuous-measurements with Markovian feedback}\label{app:conditional_master}

In this section, we unravel the weak recovery map $\tilde{\mathcal{R}}$ with Kraus operators 
\begin{equation}\label{eq:kraus_diff}
    \tilde{K}_{\ell, \pm} = \frac{1}{\sqrt{2L}}\left(I \pm \epsilon' R_\ell - \frac{\epsilon'^2}{2}R_\ell^\dagger R_\ell\right) = \tilde{U}_{\ell, \pm}\tilde{M}_{\ell, \pm},
\end{equation}
into continuous diffusive measurements with feedback that depends only on the instantaneous measurement current. This is the type of feedback given by Wiseman-Milburn \cite{wiseman, wiseman-milburn}. Specifically, we derive the \emph{conditional} master equation that corresponds to the application of this weak recovery map. From \cref{eq:kraus_diff}, we find that the positive matrices take the form
\[
\begin{split}
    \tilde{M}_{\ell, \pm} &= \sqrt{\tilde{K}_{\ell,\pm}^\dagger\tilde{K}_{\ell, \pm}} \\
    &= \frac{1}{\sqrt{2L}} \left(I \pm \epsilon' O_\ell - \frac{\epsilon'^2}{2}O_\ell^2\right) + \mathcal{O}(\epsilon'^3),
\end{split}
\]
where we have introduced the Hermitian operator
\[
    O_\ell \equiv \frac{R_\ell + R_\ell^\dagger}{2}.
\]
Similarly, applying \cref{eq:kraus_diff} to the unitary matrices yields
\[
\begin{split}
    \tilde{U}_{\ell, \pm} &= \tilde{K}_{\ell, \pm}\tilde{M}_{\ell, \pm}^{-1} \\
    &= I \mp i\epsilon' H_\ell - \frac{\epsilon'^2}{2}H_\ell^2 + i\frac{\epsilon'^2}{2}\{H_\ell, O_\ell\} + \mathcal{O}(\epsilon'^3),
\end{split}
\]
where we define a second Hermitian operator
\[
    H_\ell \equiv i\left(\frac{R_\ell - R_\ell^\dagger}{2}\right).
\]

We interpret each application of the weak recovery channel as a measurement performed by one of several observers, followed by an outcome-dependent unitary feedback. The index $\ell$ labels the observer, while $s=\pm 1$ labels the measurement outcome. Suppose that the weak channel is applied repeatedly over a coarse-grained interval $\Delta t$, with
\[
    N=\frac{\Delta t}{\delta t}.
\]
For the $j^{\mathrm{th}}$ application, let $\Lambda_j\in\{1,\ldots,L\}$ denote the selected observer and let $s_j\in\{+1,-1\}$ denote the corresponding measurement outcome. For a fixed observer $\ell$, we define the sparse
measurement record
\[
    X_{\ell,j}
    =
    s_j\,\mathbf{1}_{\{\Lambda_j=\ell\}}
    =
    \begin{cases}
        s_j, & \Lambda_j=\ell,\\
        0, & \Lambda_j\neq \ell .
    \end{cases}
\]
Conditioned on the state $\rho_c(t)$, this random variable satisfies
\[
\begin{split}
    \Pr\!\left(X_{\ell,j}=s\,\middle|\,\rho_c(t)\right)
    &=
    \operatorname{Tr}\!\left[
        \tilde{M}_{\ell,s}^{\dagger}\tilde{M}_{\ell,s}\rho_c(t)
    \right] \\
    &=
    \frac{1+2s\epsilon'\langle O_\ell\rangle_c}{2L}
    +
    \mathcal{O}(\epsilon'^3),
\end{split}
\]
where
\[
    \langle O_\ell\rangle_c
    \equiv
    \operatorname{Tr}\!\left[O_\ell\rho_c(t)\right].
\]
The accumulated measurement record associated with observer $\ell$ over the
interval $\Delta t$ is then defined by
\[
    \Delta Q_\ell(t)
    \equiv
    \frac{L\delta t}{2\epsilon'}
    \sum_{j=1}^{N}X_{\ell,j}.
\]
Using the Central Limit Theorem with $N \gg 1$ and assuming that $\rho_c(t)$ varies negligibly over the coarse-grained interval
$\Delta t$, we obtain
\[
    \Delta Q_\ell = \langle O_\ell\rangle_c\, \Delta t + \frac{1}{2\sqrt{\kappa}}\,\Delta W^{(\ell)},
\]
where $\Delta W_t^{(\ell)}$ are independent Wiener increments satisfying
\[
    \Delta W_t^{(\ell)}\Delta W_t^{(m)}=\delta_{\ell m}\,\Delta t.
\]
The measurement current can be obtained under the limit $\Delta t \rightarrow 0$ as
\[ 
    J_\ell(t) = \frac{\mathrm{d}Q_\ell}{\mathrm{d}t} = \langle O_\ell \rangle_c + \frac{1}{2\sqrt{\kappa}}\frac{\mathrm{d}W^{(\ell)}_t}{\mathrm{d}t}.
\]

For a single elementary step, conditioned on the selected observer $\ell$ and
outcome $s=\pm 1$, the normalized state update is
\[
    \rho_c(t+\delta t)
    =
    \frac{
        \tilde{K}_{\ell,s}\rho_c(t)\tilde{K}_{\ell,s}^{\dagger}
    }{
        \operatorname{Tr}\!\left[
            \tilde{K}_{\ell,s}\rho_c(t)\tilde{K}_{\ell,s}^{\dagger}
        \right]
    }.
\]
The numerator can be written as
\[
    \tilde{K}_{\ell,s}\rho\tilde{K}_{\ell,s}^{\dagger}
    =
    \frac{1}{2L}
    \bigg[
        \rho
        +
        s\epsilon'
        \left(
            \{O_\ell,\rho\}
            -
            i[H_\ell,\rho]
        \right)
        +
        \epsilon'^2
        \bigg(
            \mathcal{D}[O_\ell]\rho
            +
            \mathcal{D}[H_\ell]\rho
            -
            i[H_\ell,\{O_\ell,\rho\}]
            +
            \frac{i}{2}[\{H_\ell,O_\ell\},\rho]
        \bigg)
    \bigg]
    +
    \mathcal{O}(\epsilon'^3).
\]
The denominator is unchanged by the unitary part:
\[
    \operatorname{Tr}\!\left[
        \tilde{K}_{\ell,s}\rho\tilde{K}_{\ell,s}^{\dagger}
    \right]
    =
    \frac{1}{2L}
    \left(
        1+2s\epsilon'\langle O_\ell\rangle_c
    \right)
    +
    \mathcal{O}(\epsilon'^3).
\]
Therefore, the $j^{\text{th}}$ state update is given by
\[
    \delta\rho_j
    =
    \sum_{\ell=1}^{L}
    \left[
        \epsilon' X_{\ell,j}\mathcal{L}_{1,\ell}(\rho)
        +
        \epsilon'^2 X_{\ell,j}^2\mathcal{L}_{2,\ell}(\rho)
    \right]
    +
    \mathcal{O}(\epsilon'^3).
\]
where the superoperators $\mathcal{L}_{1,\ell}(\rho)$ and $\mathcal{L}_{2,\ell}(\rho)$ are defined as
\begin{subequations}
\begin{align}
    \mathcal{L}_{1,\ell}(\rho) &= \mathcal{H}[O_\ell]\rho - i[H_\ell,\rho], \\
    \mathcal{L}_{2,\ell}(\rho) &= \mathcal{D}[O_\ell]\rho + \mathcal{D}[H_\ell]\rho - i[H_\ell,\{O_\ell,\rho\}] + \frac{i}{2}[\{H_\ell,O_\ell\},\rho] - 2\langle O_\ell\rangle_c \left(
        \{O_\ell,\rho\} - i[H_\ell,\rho] \right) + 4\langle O_\ell\rangle_c^2\rho,
\end{align}
\end{subequations}
and the random variable $X_{\ell, j}^2$ follows
\[
    X_{\ell,j}^2 = \mathbf{1}_{\{\Lambda_j=\ell\}} =
    \begin{cases}
        1, & \Lambda_j=\ell,\\
        0, & \Lambda_j\neq \ell.
    \end{cases}
\]
By the Law of Large Numbers, we identify
\[
    \frac{1}{N}\sum_j X_{\ell, j}^2 \xrightarrow[]{N \gg 1} \frac{1}{L},
\]
giving $\epsilon'^2 \sum_j X_{\ell, j}^2 = \kappa\Delta t$. Again applying the Central Limit Theorem, we have
\[
    \epsilon' \sum_j X_{\ell, j} = 2\kappa \langle O_\ell\rangle_c\Delta t + \sqrt{\kappa}\Delta W_t.
\]
Under the limit $\Delta t \rightarrow 0$, we have the Markovian stochastic master equation corresponding to the weak recovery channel in It\^o form:
\[
    \mathrm{d}\rho = \sum_{\ell=1}^{L}
    \bigg( 
    -i[H_\ell^{(0)},\rho]\mathrm{d}t 
    -i\sqrt{\kappa}[H_\ell,\rho_c]\mathrm{d}W_t^{(\ell)} 
    -i\kappa[H_\ell,\{O_\ell,\rho\}]\mathrm{d}t
    +\kappa\mathcal{D}[O_\ell]\rho\mathrm{d}t
    +\kappa\mathcal{D}[H_\ell]\rho\mathrm{d}t
    +\sqrt{\kappa}\mathcal{H}[O_\ell]\rho \mathrm{d}W_t^{(\ell)}
    \bigg).
\]
where the base Hamiltonian $H_\ell^{(0)}$ is defined as
\[
    H_\ell^{(0)}
    =
    -\frac{\kappa}{2}\{H_\ell,O_\ell\}.
\]

To find the feedback Hamiltonian that needs to be applied, let us find the total accumulated unitary over the interval $\Delta t$:
\[
\begin{split}
    U_{\Delta t} &= \prod_\ell \prod_j \tilde{U}_{\ell, j} \\
    &= I - \sum_\ell \sum_j \bigg(i\epsilon' X_{\ell, j}H_\ell - \frac{\epsilon'^2}{2}X_{\ell, j}^2 \{H_\ell, O_\ell\}\bigg) + \mathcal{O}(\epsilon'^3) \\
    &= I - i 2\kappa \sum_\ell J_\ell(t)H_\ell \Delta t + i\frac{\kappa}{2}\sum_\ell \{H_\ell, O_\ell\}\Delta t
\end{split}
\]
Therefore, the feedback Hamiltonian be written in terms of this current as
\[
    H_{\mathrm{fb}}(t) = 2\kappa \sum_{\ell=1}^{L} J_\ell(t)H_\ell.
\]

\end{widetext}

\end{document}